\documentclass[10pt, conference, compsocconf]{IEEEtran}
\IEEEoverridecommandlockouts

\usepackage{cite}
\usepackage{amsmath,amssymb,amsfonts}
\usepackage{algorithmic}
\usepackage{graphicx}
\usepackage{textcomp}
\usepackage{xcolor}
\usepackage{multirow}
\usepackage{xspace}
\usepackage{balance}
\renewcommand\IEEEkeywordsname{Keywords}
\usepackage[labelfont={bf,small},textfont={it,small}]{caption} 
\usepackage[labelformat=simple]{subcaption}

\usepackage[numbers,sort&compress]{natbib}
\usepackage[hidelinks]{hyperref}

\usepackage{bm}

\def\BibTeX{{\rm B\kern-.05em{\sc i\kern-.025em b}\kern-.08em
    T\kern-.1667em\lower.7ex\hbox{E}\kern-.125emX}}

\usepackage{booktabs}
\newcommand{\ra}[1]{\renewcommand{\arraystretch}{#1}}

\begin{document}

\newcommand{\tian}[1]{{\color{blue}{ Tian: #1}}}
\newcommand{\suggest}[1]{{\color{red}{[#1]}}}
\newcommand{\sysname}{\textsc{Perseus}\xspace} 

\title{\sysname: Characterizing Performance and Cost \\ of Multi-Tenant Serving for CNN Models}

\author{
  \IEEEauthorblockN{Matthew LeMay}
  \IEEEauthorblockA{\textit{Worcester Polytechnic Institute}\\
  mlemay@wpi.edu}
  \and
  \IEEEauthorblockN{Shijian Li}
  \IEEEauthorblockA{\textit{Worcester Polytechnic Institute}\\
  sli8@wpi.edu}
  \and 
  \IEEEauthorblockN{Tian Guo}
  \IEEEauthorblockA{\textit{Worcester Polytechnic Institute}\\
  tian@wpi.edu}
}

\maketitle

\begin{abstract}
\label{sec:abstract}
    Deep learning models are increasingly used for end-user applications,
    supporting both novel features such as facial recognition, and traditional
    features, e.g. web search. To accommodate high inference throughput, it
    is common to host a single pre-trained Convolutional Neural Network
    (CNN) in dedicated cloud-based servers with hardware accelerators such as
    Graphics Processing Units (GPUs). However, GPUs can be orders of magnitude
    more expensive than traditional Central Processing Unit (CPU) servers. These resources could also be under-utilized facing dynamic workloads, which may result in inflated serving costs. One potential way to alleviate this
    problem is by allowing hosted models to share the underlying resources, which we
    refer to as multi-tenant inference serving. One of the key challenges is
    maximizing the resource efficiency for multi-tenant serving given hardware
    with diverse characteristics, models with unique response time Service Level
    Agreement (SLA), and dynamic inference workloads. In this paper, we present
    \sysname, a measurement framework that provides the basis for understanding
    the performance and cost trade-offs of multi-tenant model serving. We
    implemented \sysname in Python atop a popular cloud inference server called
    Nvidia TensorRT Inference Server. Leveraging \sysname, we evaluated the
    inference throughput and cost for serving various models and
    demonstrated that multi-tenant model serving led to up to 12\% cost reduction.

\end{abstract}
\begin{IEEEkeywords}
DNN inference, multi-tenancy, performance
\end{IEEEkeywords}

\section{Introduction}

Infrastructure-as-a-Service (IaaS), has emerged as a
popular option for training and deploying deep learning models, due to their
pay-as-you-go pricing models and wide selection of hardware. The increasing usage of
Convolutional Neural Network (CNN) models in computer vision applications
requires efficient utilization of cloud resources. Consequently, understanding
the cost and performance trade-offs of serving CNN model inference requests with
various cloud hardware has garnered interest from researchers~\cite{zhang2019mark,Samanta2019-wn}.

However, the typical method of serving a CNN model with dedicated resources may
lead to underutilized resources, especially when inference workloads vary.
Such inefficiency often leads to higher monetary costs; the problem becomes
more prominent when inference serving systems use expensive hardware
accelerators such as Graphics Processing Units (GPUs) for higher throughput. One
potential way to improve resource efficiency is supporting
\emph{multi-tenant inference serving}, in which models with different resource
requirements share the underlying hardware. 

As such, it is also possible to decrease serving costs by multiplexing CNN models on previously underutilized servers.

In this paper, we first show that multi-tenant model serving can achieve higher
resource utilization and lead to promising cost savings, without violating
performance guarantees for serving CNN models. Leveraging our measurement infrastructure
called \sysname, we quantified the end-user perceived latency and throughput, as
well as serving cost of running two representative CNN models on Google Cloud
Platform's Compute Engine. \sysname highlights the impacts on performance
associated with multi-tenant model serving and examines the performance and cost
tradeoffs of inference serving with different hardware configurations.

Previous literature explored the potential of using Functions-as-a-Service to achieve better resource utilization and scalability~\cite{zhang2019mark, jain2019splitserve, ishakian2018serving, tu2018pay, gunasekaran2019spock, dakkak2019trims,bhattacharjee2019barista}.
Other works have explored the use of predictive scaling~\cite{gujarati2017swayam}, Quality of Service (QoS) aware
scheduling~\cite{Qin:2019:SML:3295500.3356164, 8855453, jain2018dynamic}, GPU
primitive sharing~\cite{yu2019salus}, and edge-based
techniques~\cite{ogden2018modi, li2018auto, ogden2020mdinference, gilman2019challenges} to improve serving efficiency. Our work complements prior research by
providing the basis for understanding the performance implications and for
improving resource utilization of cloud-based inference servings.
We make the following key contributions.

\begin{itemize}
    \item Our study demonstrates the need for 
    multi-tenant model serving, and shows up to 12\% cost savings when
    appropriately mixing inference workloads.
    \item We designed and implemented a suite of tools, collectively referred to as
    \sysname
    \footnote{https://github.com/cake-lab/perseus},
    that facilitates further evaluation of performance and cost trade-offs for new model serving scenarios, such as running new
    CNN models on different GPUs.
    \item We identify a number of aspects, including inefficient framework
    supports for CPU inference and for model caching, that hinder the observed
    inference performance. Our findings shed light on and pave the way for
    complementary research such as resource provisioning and load balancing for
    model serving. 
\end{itemize}

The remainder of this paper is structured as follows:
Section~\ref{sec:background} introduces the key concepts underpinning CNN model
serving systems and discusses related work.
Section~\ref{sec:problem_methodology} presents the problem statement followed by
the design of \sysname and our measurement methodology for characterizing
multi-tenant model serving, as presented in Section~\ref{sec:eval}. 
Finally, we summarize the findings of our research and potential directions in Section~\ref{sec:conclusion}.

\section{Background and Related Work}
\label{sec:background}

There are numerous existing frameworks~\cite{zhang2019mark,
crankshaw2017clipper, romero2019infaas, olston2017tensorflow, tensorrt,
predictionio, redisai, bhattacharjee2019barista} and services~\cite{aiplatform,
azureml, sagemaker} for supporting inference serving in cloud environments. We
briefly describe these inference systems and common deployment practices. Then
we discuss the hardware in which inference serving platforms leverage and
holistic techniques for evaluating inference serving systems. 

\begin{table}[t]
\sffamily
\ra{1.3}
\begin{center}
\resizebox{0.42\textwidth}{!}{
\begin{tabular}{rrrr}
\toprule
\textbf{GPU Type}        & \textbf{Memory (GB)} & \textbf{Memory Bandwidth} & \textbf{Cuda Cores} \\ \midrule
Nvidia Tesla P4 & 8               & 192 GB/s                     & 2560                   \\ \hline
Nvidia Tesla V4 & 16              & 320 GB/s                     & 2560                   \\ \bottomrule
\end{tabular}
}
\end{center}
\caption{Overview of evaluated Nvidia GPU devices.}
\label{table:specs}
\end{table}

\subsection{Inference Serving Frameworks}
\label{sec:inference_serving_framework}

Inference serving frameworks have evolved to support a wide array of use cases,
libraries, and platforms. TensorFlow Serving~\cite{olston2017tensorflow} is one
of the initial open-source inference serving systems that leverages GPUs.
TensorFlow Serving also supports multi-model deployments and provides an endpoint for
prediction, but requires models to be trained using TensorFlow explicitly. Other frameworks such as PredictionIO and
RedisAI~\cite{predictionio,redisai} allow the serving of models trained using
different frameworks. Further, frameworks such as Nvidia's TensorRT Inference
Server~\cite{tensorrt} provide hardware-specific inference optimizations, e.g.,
for Nvidia's GPUs.

Several frameworks have evolved to incorporate additional features aiming to
improve performance of inference serving. Clipper~\cite{crankshaw2017clipper}
adds additional functionality to ensure SLAs and to achieve better prediction
accuracy. MArk~\cite{zhang2019mark} and Barista~\cite{bhattacharjee2019barista}
leverage Functions-as-a-Service (FaaS) to handle and scale transient workloads in order to maintain SLAs. INFaaS~\cite{romero2019infaas} shares models
and hardware across applications by optimizing model deployment and autoscaling
mechanisms. However, INFaaS focuses on a single VM configuration of either CPU
and GPU, and uses GPU memory constraints to scale each model serving independently. This can lead to resource
under-utilization especially when models serve dynamic inference requests. In
contrast, we explore the inference cost savings of \emph{sharing resources
without constraints} through evaluating resource footprints of different
model-hardware configurations.

\subsection{Inference Serving Hardware}
\label{sec:inference_serving_hardware}

The abundance of commodity CPU servers in the cloud makes them ideal candidates
for serving inference requests~\cite{Liu2019-sf,Soifer2019-ln, Zhang2019-lp},
while the emergence of hardware accelerators provide new opportunities
and challenges. Among the plethora of accelerators, GPUs have become the most
popular type and are closely associated with deep learning. Manufacturers have
been making highly specialized GPUs for different deep learning tasks, such as 
\emph{Nvidia P4 GPU} for inference jobs. Table~\ref{table:specs} shows hardware
specifications of two such GPUs. In this paper, we
chose to focus on GPU inference for three reasons. \textit{First,} GPUs are
widely used in deep learning, particularly in the cloud environment.
\textit{Second,} GPUs exhibit intricate advantages and shortcomings compared to
CPUs. For example, GPUs are ideal for highly parallel computation such as matrix
multiplication which dominates CNN inference, while their performance are
fundamentally constrained by limited GPU memory and slower memory transfer
between CPU and GPU. \textit{Third,} cloud-based GPUs are much more
expensive, leading to large room for improvements of monetary cost.

\subsection{Inference Serving Deployments}
\label{sec:inference_serving_deployments}

Deploying a CNN model to a pre-provisioned server requires developers to adhere
to a given framework's workflow. Namely, pre-trained models, with their weights
and labels, must be exported into a format supported by the serving framework.
Inference servers commonly expose endpoints such as REST, gRPC, or client API
interface, which can be used to query a model~\cite{zhang2019mark,
romero2019infaas, dakkak2019trims, crankshaw2017clipper,
bhattacharjee2019barista, azureml, aiplatform, sagemaker, predictionio, redisai,
tensorrt, olston2017tensorflow}. In systems that support autoscaling
\cite{zhang2019mark, romero2019infaas, dakkak2019trims, crankshaw2017clipper,
bhattacharjee2019barista}, middleware manages provisioning and acts as a single
endpoint which routes requests to individual model serving.

Several major cloud providers offer managed inference serving frameworks such as
Amazon's SageMaker, for deploying a single model in an isolated
environment~\cite{azureml,aiplatform,sagemaker}. These services abstract
the deployment process and provide high-level tools for autoscaling individual
models. Amazon's Elastic Inference introduced the ability
to acquire and attach a portion of a GPU's resource to a SageMaker
instance~\cite{elasticinference}, further reducing over-provisioning.

\begin{table}[t]
\sffamily
\ra{1.3}
\begin{center}
\resizebox{0.41\textwidth}{!}{
\begin{tabular}{ccrr}
    \toprule 
\textbf{Machine Type}  & \textbf{GPU Type}        & \textbf{GPU Count} & \textbf{Cost (\$/hour)} \\ \hline
n1-standard-8 & Nvidia Tesla P4 & 1              & 0.688           \\ \hline
n1-standard-8 & Nvidia Tesla P4 & 2              & 1.108           \\ \hline
n1-standard-8 & Nvidia Tesla T4 & 1              & 0.933           \\ \hline
n1-standard-8 & Nvidia Tesla T4 & 2              & 1.598           \\ \hline
n1-standard-8 & N/A & N/A              & 0.268           \\ \bottomrule
\end{tabular}
}
\end{center}
\caption{Cloud server unit cost with GPU configurations from Google Cloud \texttt{us-central1-a}, as of November 26th, 2019.}
\label{table:prices}
\end{table}

\subsection{Inference Serving Performance Characterization}

There are a plethora of choices when deploying inference serving systems;
therefore, it is important to determine a model's characteristics for a
given framework in a specific system. The first-order goal of inference serving
is latency. Adhering to latency SLAs is one of the key challenges of inference
serving, especially for applications that require real-time performance.
Consequently, latency determines the viability of performing inference with a
given configuration. SLA compliance is commonly measured by verifying that the
$95^{th}$ or $99^{th}$ percentile of the end-to-end response time of recent
requests is below a predefined threshold~\cite{zhang2019mark,
gujarati2017swayam}. The second-order goals of inference serving are throughput
and cost. Deployments can require handling a large number of requests in a short time frame~\cite{Soifer2019-ln}, thus accurately evaluating the inference throughput can help determine performance bottlenecks under heavy loads.
Throughput is commonly measured by estimating the peak or steady-state request
rate of the system~\cite{crankshaw2017clipper, reddi2019mlperf}.
Table~\ref{table:prices} shows the unit cost of performing inference on
chosen cloud server configurations.

\section{Problem Statement and Measurement Methodology}
\label{sec:problem_methodology}

\subsection{Problem Statement}
\label{sec:problem_statement}

In this paper, we investigate the performance and cost trade-offs of
multi-tenant model serving compared to single-tenant serving as well as CPU serving. Such understandings can improve the resource
efficiency of serving CNN models using cloud servers of various capacities. 
To do so, we designed and implemented a measurement framework called
\sysname which we then leveraged to quantify the model serving performance of
various configurations. The configurations we explored include serving inference
requests with CPU-only vs. with GPU hardware accelerator, as well as dedicated
vs. shared GPU resources. Our measurements pinpoint several potential
performance bottlenecks when serving CNN inference requests and demonstrate the
cost savings prospect of GPU-based multi-tenant model serving.

\subsection{\sysname Architecture}

\begin{figure}[t]
    \centering
    \includegraphics[width=0.45\textwidth]{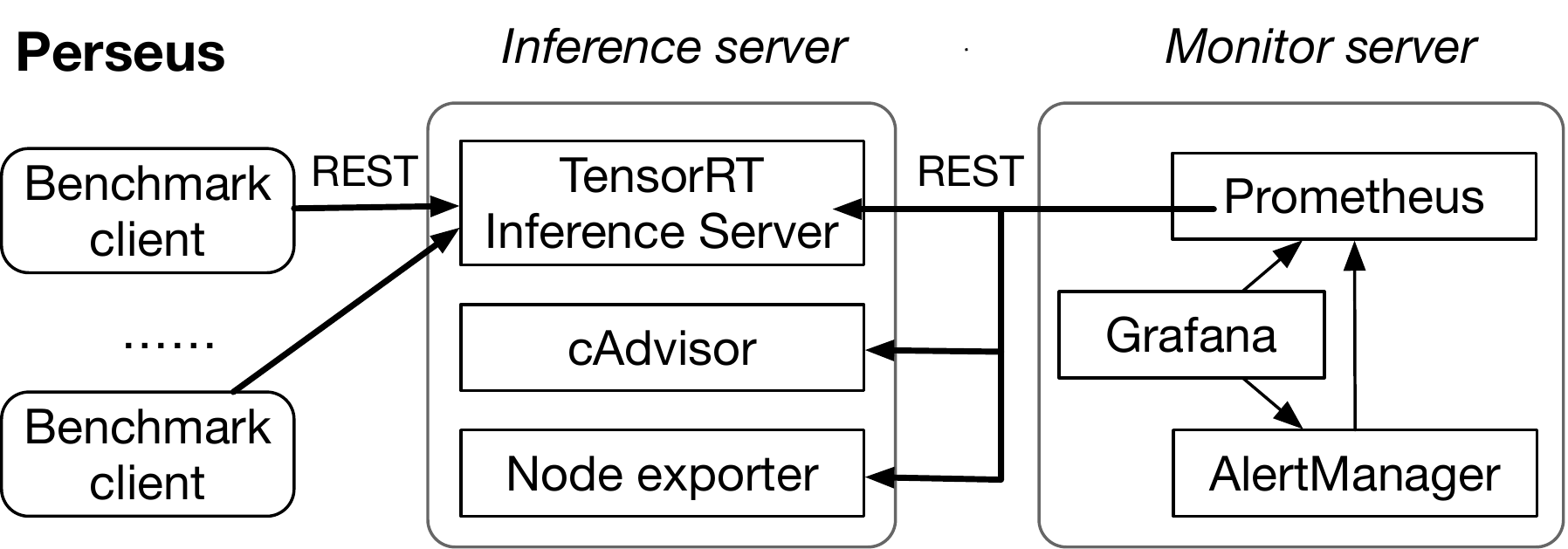}
    \caption{Architecture of our measurement framework \sysname.}
    \label{fig:architecture_v2}
\end{figure}

In the past, there have been many works outlining the general procedures for
determining the performance characteristics of an application on a server~\cite{demirkan2013leveraging}. Due
to the domain-specific intricacies of inference serving and application-specific
constraints of working with an existing framework such as Nvidia's TensorRT
Inference Server~\cite{tensorrt}, we therefore propose a new measurement framework called \sysname.
The framework serves the purpose of gathering accurate performance data
relevant to the server and models being served. We first describe the design and
implementation of \sysname and then outline the models evaluated, the workloads
used, and the experimental setup using the aforementioned framework.

Figure~\ref{fig:architecture_v2} shows the implementation of \sysname. All components are encapsulated in Docker containers
to ensure reproducibility. \sysname currently supports benchmarking image
classification applications with Convolutional Neural Networks.
The benchmarking client preprocesses the input data from a given directory,
stores the preprocessed data in the client's memory, and then generates and
sends inference requests to the inference server. The number of clients in
conjunction with its computation capacity determines the peak throughput
($\lambda$) in inference requests per second for a given model. The client is
able to achieve an accurate estimation without using server-side statistics. The
inference server is based on Nvidia's TensorRT Inference Server~\cite{tensorrt}
and includes two additional components \emph{cAdvisor} and \emph{Node exporter}
that aggregate and export the performance characteristics and resource
utilization data such as GPU, CPU and network utilization of running
containers. We further use \emph{Prometheus} in conjunction with
\emph{AlertManager} to store collected data and \emph{Grafana} to visualize
performance data.

\subsection{Measurement Methodology}
\label{sec:measurement_methodology}

\subsubsection{Experimental Testbed}
We used \emph{n1-standard-8} instances on Google Compute Engine, with 8 Intel Broadwell vCPUs and 30GB of RAM, as
the platform for each client and server. Each instance ran a minimal
installation of Ubuntu 18.04.3 LTS using Linux 5.0.0-1021-gcp as the 64-bit
kernel. We used Docker version 19.03.4 and Containerd version 1.2.10 hosting Nvidia's
TensorRT Inference Server version 1.6.0. cAdvisor version 0.33.0 and Node
exporter version 0.18.1 were used to collect the server's resource and
performance information. We use version 1.6.0 Nvidia's TensorRT Inference
Framework Python Client SDK to perform inference requests using Python 3.6.8 on
each client. The monitoring stack was composed of Prometheus version 2.11.1
and Grafana version 6.3.3. We chose to evaluate the GPU inference performance using \emph{Nvidia's P4} and \emph{T4 GPUs} due to their wide adoption, low price-point, and designation as data center inference products \cite{nvidiainferenceplatforms}.

\subsubsection{Model Selection}
We used two popular CNN models, \textit{Inception-V3}~\cite{szegedy2016rethinking} and
\textit{ResNet50}~\cite{he2016deep}, as the basis for evaluating the performance
characteristics of inference. The models, which perform image classification tasks,
require an image as input and produce a string as output. These two models were
implemented in different frameworks: \textit{Inception-V3} uses
TensorFlow~\cite{tensorflow} and \textit{ResNet50} uses Caffe2~\cite{caffe2}. This allowed
for the use of original unmodified, pre-trained models and guarantees isolated
model runtimes. It also demonstrated the framework-agnostic approach of \sysname.

\subsubsection{Workloads}
In our experiment we opted for a dataset of 6908 images, which was a randomly selected subset of the Open Images V3 Validation Dataset \cite{openimages}. The images were preprocessed before inference to eliminate the overhead of loading and processing images at runtime. \textit{Inception-V3} requires 299 by 299 pixels RGB images and \textit{ResNet50} 224 by 224 pixels RGB images. The dataset's size provides the advantage of reducing the effects of abnormalities on results while maintaining a short runtime. Our framework delegates batching to the server where the server could treat each request as a single request from a client. The batch size determines the latency and throughput of the server. Therefore, we used the same batch size across all hardware configurations to provide a fair comparison.

\subsubsection{Metrics} We evaluated the efficacy of the cost and performance tradeoff of multi-tenant model serving using \sysname framework, models, and workloads. We conducted several experiments to show the effect of hosting an additional model on startup time, latency, and throughput. \emph{Our key goal is to understand whether overhead introduced during multi-tenancy significantly impacts performance of inference.} Peak or steady-state throughput \emph{$\lambda$} measures the maximum rate of inference requests over a time span. The latency of requests \emph{t} denotes the end-to-end response time for an inference request, where the $95^{th}$ percentile latency is the latency for 95\% of requests. \textit{Cost per one million inference requests} \emph{c} provides a standardized metric which promotes price comparisons across server hardware by accounting for the relative performance of a device~\cite{crankshaw2017clipper}. 

More broadly, the measurement statistics are utilized to estimate the performance metrics for each stage of inference serving in regard to the hardware used. The startup time of various hardware platforms measures the time required to begin loading a model for inference on pre-provisioned server instances. The \textit{dedicated model performance} is calculated by measuring the maximum computation capacity of a server under peak throughput \emph{$\lambda$}, thus determining the stable operating range of a model on a given configuration. Finally, the \textit{multi-model performance} is determined by measuring the resulting latency and throughput of each model served. The overhead and tradeoff of hosting multiple models is conveyed through the shift in peak workload performances and each model's performance relative to its counterparts.

\section{Performance and Cost Characterization}
\label{sec:eval}

Utilizing \sysname, we evaluated on various serving options for deep learning
inference. In practice, CNN models have been widely deployed and served with CPU
only~\cite{Liu2019-sf,Soifer2019-ln, Zhang2019-lp}. In this section, we first
evaluate and study the tradeoffs of inference using CPUs versus GPUs. We then
characterize the benefit of multi-tenant model serving with GPUs by comparing against dedicated GPU inference. 

\subsection{CPU vs. GPU inference}
\label{sec:cpu_vs_gpu}

\begin{figure}[t]
    \centerline{\includegraphics[width=0.35\textwidth]{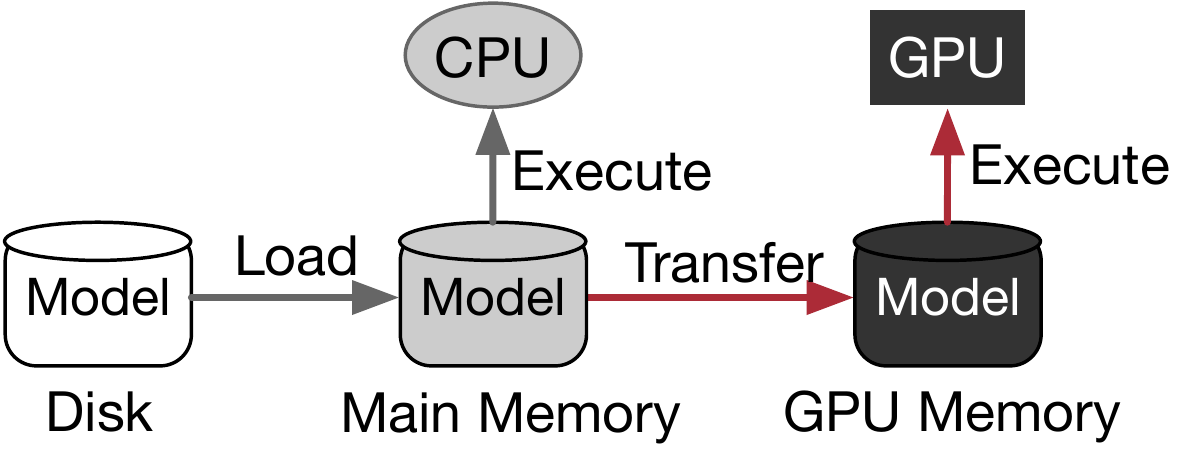}}
    \caption{Measurement illustration for CPU vs. GPU inferences.}
    \label{fig:cpu_gpu_measurement_setup}
\end{figure}

\begin{table}[t]
    \sffamily
    \ra{1.3}
    \resizebox{0.48\textwidth}{!}{
    \begin{tabular}{@{}llrclrclrcl@{}}
     \toprule
      && \multicolumn{9}{c}{Time to Execute on Device (ms)}\\
      \cmidrule{3-11}
      && \multicolumn{3}{c}{\textbf{n1-standard-8 CPU}} &
      \multicolumn{3}{c}{\textbf{Nvidia P4 GPU}} &
      \multicolumn{3}{c}{\textbf{Nvidia T4 GPU}} \\ \midrule
      \textbf{ResNet50} & Hit     & 159.1 &$\pm$& 3.4
      & 18.5 &$\pm$& 0.6                         & 18.2 &$\pm$& 0.1                         \\
      & Miss    & 1401.4 &$\pm$& 89.9                          & 18418.4
      &$\pm$& 498.5                     & 21264.4 &$\pm$& 310.6
      \\ 
      \textbf{Inception-V3} & Hit & 75.1 &$\pm$& 1.2
      & 217.7 &$\pm$& 1.8                        & 325.9 &$\pm$& 7.3                        \\
      & Miss  & 3806.7 &$\pm$& 222.5                         & 18704.8 &$\pm$&
      343.4                    & 21693.3 &$\pm$& 763.7                    \\
      \bottomrule
     \end{tabular}
    }
    \caption{Average time ($t$) to perform inference for CPU versus 
    GPU hardware. A hit means the model was already present in main memory (when
    executing on the CPU) or in GPU memory (when executing on the GPU). A miss
    requires either loading models from the disk into the main memory or from
    the disk to GPU memory.
    }
     \label{table:cpu_gpu_time}
\end{table}

We first quantify the inference performance of two popular Convolutional Neural
Networks with comparable model sizes, number of parameters, and inference
accuracy (i.e., \textit{InceptionV3} and
\textit{ResNet50}) to demonstrate the importance and challenges
of determining the appropriate serving hardware given the workload i.e. CNN models. Table~\ref{table:cpu_gpu_time} summarizes the
inference time (batch size of 1) when executed on the CPU or a discrete GPU.
When executed on the CPU, we define a \emph{hit} to mean that the model was
already present in main memory (i.e., RAM) and a \emph{miss} to mean that the
model first needed to be loaded from disk. When executed on one of the two GPUs,
we define a \emph{hit} to mean that the model was already present in \emph{GPU
memory} and a \emph{miss} to mean that the model needed to be loaded from disk
into main memory and then transferred to GPU memory. Our measurement was
conducted using Google Cloud, following the setup in
Figure~\ref{fig:cpu_gpu_measurement_setup} and leverages our \sysname
measurement infrastructure (Figure~\ref{fig:architecture_v2}).

We make three key observations. \emph{First,} static model characteristics, such
as model file size, are not a good indicator of runtime requirements and
performance. \emph{Second,} it is not always faster to execute the model on a
GPU, even with GPUs optimized for inference such as \emph{NVIDIA's P4} and
\emph{T4 GPUs}. For example, in the case of \textit{Inception-V3} (hit),
it is more than three times faster to execute using an Intel CPU than the
\emph{T4 GPU}.  However, we measured the peak throughput of both GPUs to be 12 times higher than that of the CPU with a
batch size of 8. \emph{Third,} even
though the on-disk sizes of these two models are roughly the same, it takes twice
as long to load \textit{Inception-V3} into CPU memory but nearly the same amount
of time to transfer each model from CPU to GPU memory. Our measurements both
demonstrate the intricate trade-offs between caching in CPU memory versus GPU
memory and motivate the need to mask the data transfer latency to the GPU
memory.

\begin{table}[t]
\sffamily
\ra{1.3}
\begin{center}
\resizebox{0.34\textwidth}{!}{
\begin{tabular}{rrrr}
\toprule
  & \multicolumn{1}{c}{\textbf{c (\$)}}
  & \multicolumn{1}{c}{\textbf{$\bm{t_{95}}$ (sec)}} &
  \multicolumn{1}{c}{\textbf{$\bm{\lambda}$ (reqs/sec)}} \\ \hline
\textbf{ResNet50} & 16.836 & 2.473              & 4.724           \\ \hline
\textbf{Inception-V3} & 4.029 & 0.765              & 19.720           \\ \bottomrule
\end{tabular}
}
\end{center}
\caption{Inference performance and cost with n1-standard-8.
We used a batch size of 8 and measured the cost $c$, $95^{th}$ percentile
latency and peak throughput $\lambda$ when serving 1 million requests. 
}
\label{table:cpuinference}
\end{table}

Table~\ref{table:cpuinference} shows the performance evaluation of CPU-based
inference using same hyper-parameters as the GPU inference (i.e. a single model
instance with a batch size of 8). Under steady-state conditions, the cost of
performing one million inference requests at peak throughput on an 8-core CPU is significantly higher than GPU based inference under peak throughput conditions shown in Table~\ref{table:singlep4perf}.
The much worse performance of \textit{ResNet50} when serving batched requests is
largely due to inference framework's limited support for CPU inference.
Specifically we observe that Caffe2 accumulated and processed batched requests
on a single core. This suggests the need to carefully choose inference
frameworks that are optimized for the underlying hardware~\cite{Hazelwood2018-rk,Park2018-li,Liu2019-sf}.
Therefore, while CPU-based inference may be able to swiftly adapt to transient
load spikes, it is not a cost and performance effective solution for handling
workloads demanding higher throughput.

\textbf{Summary:} Serving with a cold cache is always better on CPU servers due to the high data
transfer latency between CPU memory and GPU memory. Inference model miss incurs
a time cost overhead ranging between 67X to 1168X compared to model hit on GPUs.
While on CPUs, the overhead of model miss is at most 51X. However, some models
are better suited for GPU serving with warm cache. For example,
\textit{ResNet50} model on hit is up to 14X faster on GPU than on CPU, which
does not hold true for \textit{InceptionV3}.

\subsection{CNN inference on GPU}

\subsubsection{Characterization of Dedicated Model Inference on GPUs}

We profiled each model on available hardware configurations to establish the baseline performance for GPU based inference. 
Table~\ref{table:singlep4perf} shows the dedicated model inference serving results using different GPU types and counts.
Across both GPU types, the cost per inference and latency decrease when the number of GPUs increases. 
Accordingly, two GPU instances achieved higher overall throughput compared to the single GPU instances.
On average, by adding an additional GPU, the price per inference request decreased by 14.43\% for \textit{ResNet50} and 14.00\% for
\textit{Inception-V3}. 
Surprisingly, for both CNNs, increases in peak throughput \textit{$\lambda$} generated by \emph{T4 GPUs} yielded a higher cost. 
Furthermore, running on a single \emph{P4 GPU} \textit{Inception-V3} achieved a peak GPU utilization of 92\% compared to ResNet50's utilization of 88\%.
More importantly, on the GPU memory, utilization of \textit{Inception-V3} was 97.20\% compared to 21.58\% for ResNet50. In scenarios where maximum utilization can be achieved, the Dual \emph{P4 GPU} configuration achieves the best cost-performance outcome.

\begin{table*}[t]
\sffamily
\ra{1.3}
\begin{center}
\resizebox{0.9\textwidth}{!}{
\begin{tabular}{crrrrrrrrrrrr}
\toprule 
                                   & \multicolumn{3}{c}{\textbf{One P4 GPU}}
                                   & \multicolumn{3}{c}{\textbf{Two P4 GPUs}}
                                   & \multicolumn{3}{c}{\textbf{One T4 GPU}}
                                   & \multicolumn{3}{c}{\textbf{Two T4 GPUs}}
                                   \\ \cmidrule{2-13} 
                                   & \multicolumn{1}{c}{\textbf{c (\$)}}
                                   & \multicolumn{1}{c}{\textbf{$\bm{t_{95}}$ (sec)}} &
                                   \multicolumn{1}{c}{\textbf{$\bm{\lambda}$ (reqs/sec)}} &
                                   \multicolumn{1}{c}{\textbf{c (\$)}} &
                                   \multicolumn{1}{c}{\textbf{$\bm{t_{95}}$ (sec)}} &
                                   \multicolumn{1}{c}{\textbf{$\bm{\lambda}$ (reqs/sec)}} &
                                   \multicolumn{1}{c}{\textbf{c (\$)}} &
                                   \multicolumn{1}{c}{\textbf{$\bm{t_{95}}$ (sec)}} &
                                   \multicolumn{1}{c}{\textbf{$\bm{\lambda}$
                                   (reqs/sec)}} &
                                   \multicolumn{1}{c}{\textbf{c (\$)}} &
                                   \multicolumn{1}{c}{\textbf{$\bm{t_{95}}$ (sec)}} &
                                   \multicolumn{1}{c}{\textbf{$\bm{\lambda}$ (reqs/sec)}}
                                   \\
                                   \hline
\multicolumn{1}{c}{\textbf{ResNet50}}     & 0.938 & 0.076
& 203.810                           & 0.773                      & 0.059
& 398.150                           
& 1.012 & 0.077                      & 256.088
& 0.898                      & 0.048                      & 494.608

\\ \hline
\multicolumn{1}{c}{\textbf{Inception-V3}} & 1.235 & 0.102
& 154.801                           & 1.008                      & 0.080
& 305.199                          
& 1.249 & 0.061
& 207.44                           & 1.129                      & 0.059
& 393.311                  
\\ \bottomrule
\end{tabular}
}
\end{center}
\caption{Inference performance and cost with P4 and T4 GPUs.
Serving with two P4 GPUs can be 18.4\% cheaper for 1 million requests due to CPU
cost amortization and linear throughput scalability. Increasing the number of T4 GPUs from 1 to 2, decreases the cost and latency of inference and increase the peak throughput across both models, results in 37.7\% cost saving.
}
\label{table:singlep4perf}
\end{table*}

\begin{figure}[t]
    \centerline{\includegraphics[width=0.42\textwidth]{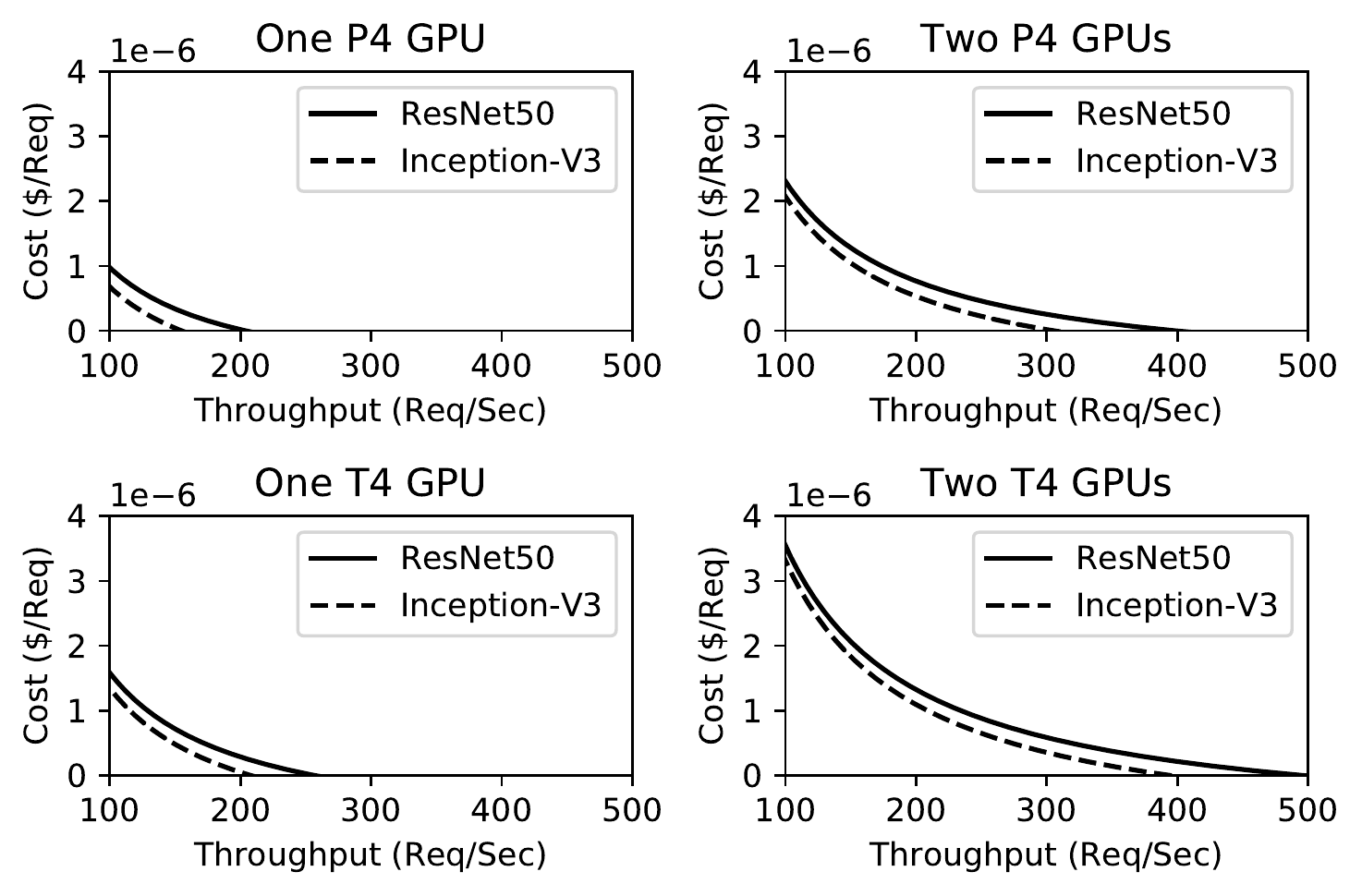}}
    \caption{The effects of throughput on inference cost showing the polynomial increase in cost across all hardware configurations. 
    }
    \label{fig:singlecosts}
\end{figure}

Our data shows that under peak loads, GPU resources are under-utilized, i.e. cannot be fully leveraged, in multi-tenant inference. This suggests that performing multi-tenant inference on \emph{P4 GPU} will lead to over-utilized GPU memory, due to the large size of each model in the card's memory.
As shown in Figure \ref{fig:singlecosts}, in conditions where the peak throughput is not met, the cost increases polynomially as the throughput decreases.
The results show that in scenarios of under-utilization, optimizing the hardware cost requires accurately estimating a model's throughput.
This in turn calls for understanding and modeling the characteristics of CNN models on how under-utilization can be effectively mitigated when serving another model to utilize the remaining resources. 

While our experiment results suggest that testing a server instance with four
GPUs may produce additional cost savings, we encountered issues when testing
this configuration. When the four P4 or T4 GPUs were configured, the system
became unstable. Subsequently, the data collected for the peak workload
\textit{$\lambda$} and latency \textit{t} did not prove reliable. Our tests
determined that the peak throughput for ResNet50 on four P4 GPUs achieved
between 110\% and 130\% above the peak throughput of the equivalent two GPU
servers. Ignoring the variability, the price per request at the maximum
throughput did not justify the price of the server. In our experiments, the CPU,
RAM, and GPU utilizations did not pose as the bottleneck. Additionally, the
experimental findings of a Google Cloud Platform Blog article~\cite{gcpperf},
which showed the average throughput of the network to be 10 GB/s, supports the
assertion that our workloads of 0.5-1.0 GB/s did not result in a network
bottleneck. The bottleneck was experimentally determined to be server's gRPC
endpoint.

\textbf{Summary:} For our selected workload on dedicated GPU server for inference, it is better to have more than one GPU card to share the workload, in terms of cost per request, when the request rate is higher than 100 requests/sec. Furthermore, even with a single GPU card, the GPU memory utilization for some models is underutilized, which could potentially be improved by serving multiple models on the same server.

\subsubsection{Multi-Tenant Model Serving}

\begin{figure}[t]
    \centering
    \includegraphics[width=0.4\textwidth]{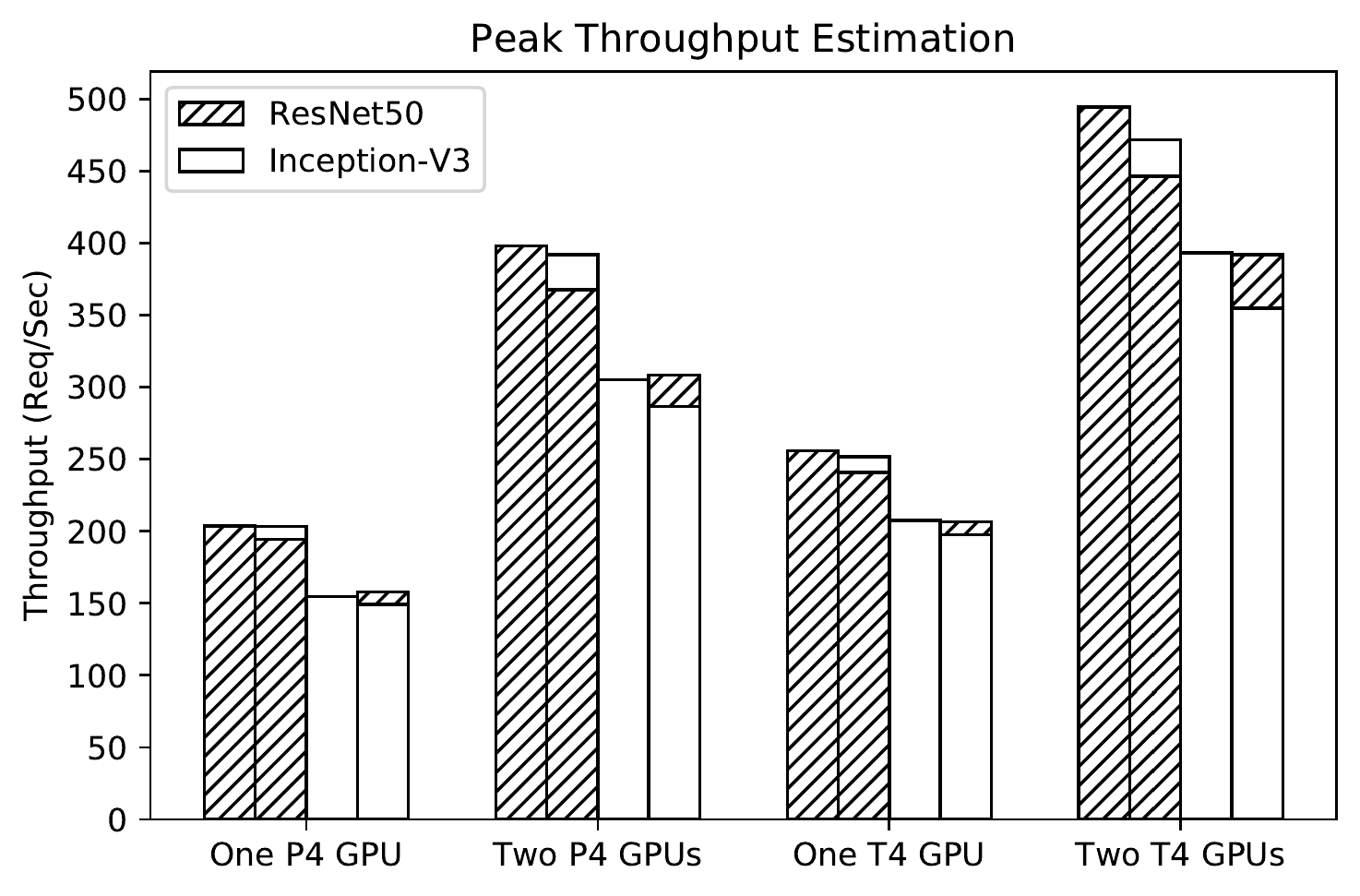}
    \caption{Comparison of peak inference throughput for single vs. multi-tenant model serving. The first and third bars of each group represent the single dedicate model serving throughput, while the second and fourth bars describe the multi-tenant counterparts.}
    \label{fig:multi_peak_throughput}
\end{figure}

To quantify the benefit of multi-tenant model serving, we evaluated the peak throughput, $95^{th}$ percentile latency, and serving costs when the two CNN models share the underlying resources.
Figure~\ref{fig:multi_peak_throughput} compares the achieved peak inference throughputs of \textit{ResNet50}-dominated (and \textit{Inception-V3}-dominated) requests sharing underlying GPU(s) with \textit{Inception-V3} (and \textit{ResNet50}) to those
of serving these two CNNs on the corresponding
dedicated GPU(s), respectively.
We observe that a multi-tenant model serving with such extreme workload mixes (e.g., 1:20 ratio) can achieve comparable throughput
to a dedicated single model serving with one GPU.
However, in the case of two GPUs, the aggregate throughput of multi-tenant model serving slightly lagged behind. 

\begin{figure}[t]
    \centering
    \includegraphics[width=0.45\textwidth]{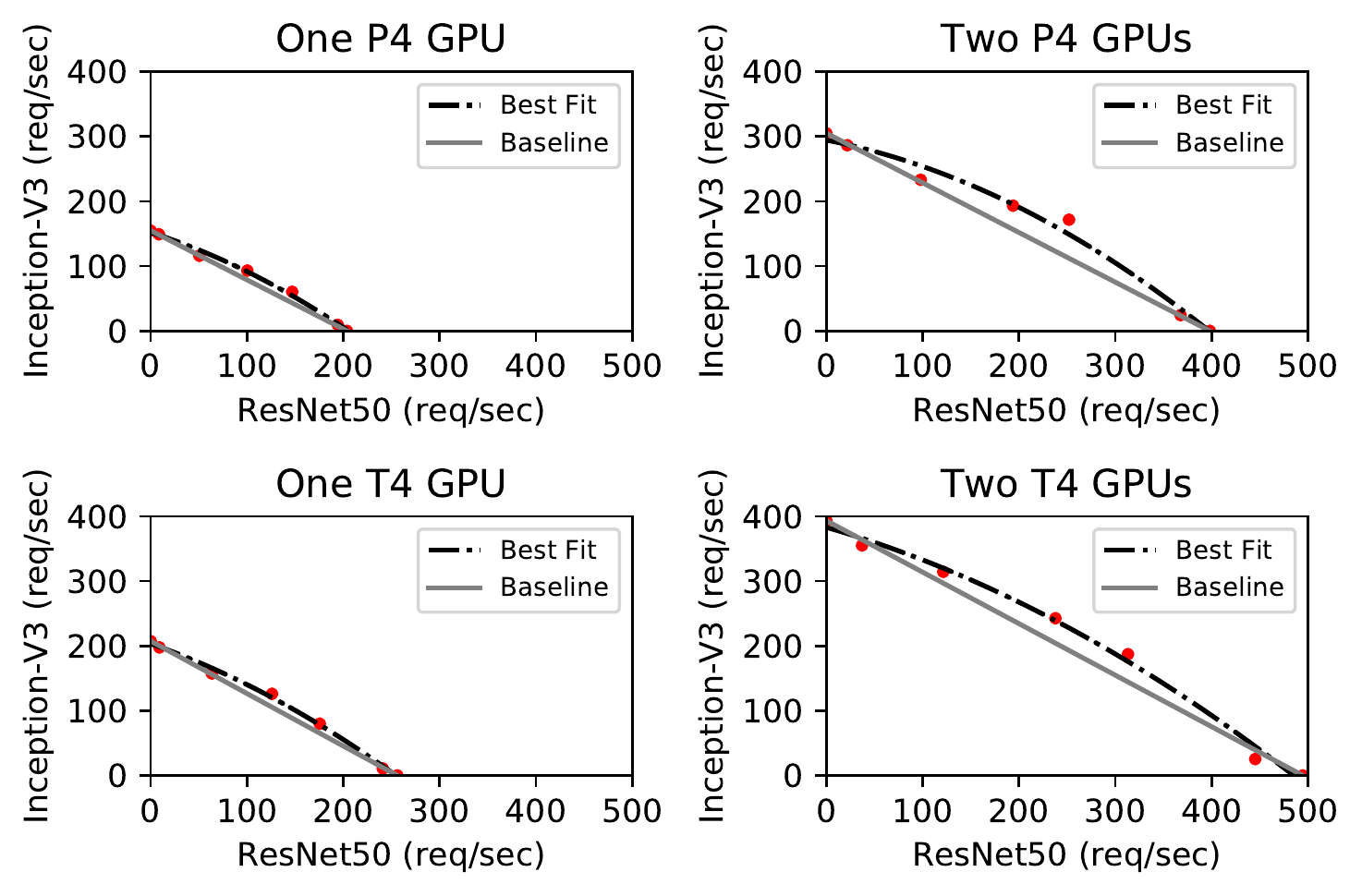}
    \caption{Inference serving throughputs of multi-tenant model serving with
    different workload mix ratios.
        }
    \label{fig:multithroughputgraph}
\end{figure}

To understand the interplay between different multi-tenant inference workloads, we repeated the above measurements by adjusting the ratio of model requests. 
Figure~\ref{fig:multithroughputgraph} shows the achieved throughputs of \textit{ResNet50} and \textit{Inception-V3}. 
The results show that it is a non-linear relationship as the requests for two models change.
Thus, the overhead of hosting an additional model is less than the performance gain of exploiting under-utilized resources.
This means hosting two models that cannot be both fully loaded onto a GPU's memory does not make multi-tenant inference impractical.
The performance gain occurs when the throughput is consistently achieved and both models are experiencing non-trivial workloads (i.e. between 25\% and 75\% of their peak throughput).
Furthermore, we observe that the cost per inference and latency decrease when the number of GPUs increases, for both GPU types.

To quantify the relative cost saving of multi-tenant model serving with different workload mix ratios, we define a metric called \emph{effective unit cost}.
For a given server that costs $a$ dollars per hour, if its capacity for serving \textit{ResNet50} is $x$ requests per hour and for serving
\textit{Inception-V3} is $y$ requests per hour, then we can derive the
server-model unit cost as $\frac{a}{x}$ and $\frac{a}{y}$. The \emph{effective
unit cost} is defined as $b= \frac{ax'}{x} + \frac{ay'}{y}$ where x' and y' are
the number of requests the server can dedicatedly serve \textit{ResNet50} and
\textit{Inception-V3}, respectively.
Intuitively, $a$ is the actual cost using multi-tenant model serving, and $b$ describes how much one needs to pay for serving an aggregate request rate of $x'+y'$.
Therefore, the cost saving of multi-tenant model serving can be calculated as $\frac{(b-a)}{a}$.

\begin{table}[t]
\sffamily
\ra{1.3}
\begin{center}
\resizebox{0.42\textwidth}{!}{
\begin{tabular}{lrrrr}
\toprule
  & \textbf{One P4 GPU} & \textbf{Two P4 GPUs} & \textbf{One T4 GPU} & \textbf{Two T4 GPUs} \\
\midrule
$\bm{a}$ \textbf{ (\$/hour)} & 0.688      & 1.108       & 0.933      & 1.598   
\vspace{2pt} \\ 
$\bm{b}$ \textbf{ (\$/hour)} & 0.753      & 1.241       & 1.026      & 1.754    \\
$\textbf{Savings (\%)}$ & 9.45\% & 12.00\% & 9.96\% & 9.76\% \\ 
\bottomrule
\end{tabular}
}
\end{center}
\caption{Comparison of the lowest effective unit cost of multi-tenant model
serving to server unit cost. 
}
\label{table:table_multi_tenant_comparison}
\end{table}

Table~\ref{table:table_multi_tenant_comparison} shows the results of performing the aforementioned calculations. When the request rates for both models converge to the same request rate, the effective unit cost is higher than that of a server hosting a single model.
We show that across all hardware GPU configurations, it is on average 9.5\% cheaper to serve the two models in this configuration.
The steady-state latency for \textit{ResNet50} and \textit{Inception-V3} increased by 55\% and 26\% respectively.
The results revealed that the best cost-performance ratio is achieved when both models are serving at the same request rate.
In addition, serving multiple models can effectively achieve higher utilization of resources when a single model server experiences under-utilization.

\textbf{Summary:} Multi-tenant model serving can reduce the effective unit cost by up to 12\% with \emph{two P4 GPUs}.
The maximum cost reduction for each hardware configuration was achieved when serving \textit{ResNet50} and \textit{Inception-V3} at roughly the same throughput.
Our observations further suggest the benefits of intelligent provisioning and scheduling of inference requests using a multi-tenant approach.

\section{Conclusion and Future Work}
\label{sec:conclusion}

As pre-trained deep learning models have been increasingly utilized for new
application features and integrated into existing applications, it necessitates
the research of resource-efficient inference serving. In this paper, we
demonstrated the benefits of multi-tenant model serving, a promising way to
improve server resource utilization and reduce monetary cost. We quantified its achieved performance and
cost by comparing to other common serving configurations, using our measurement framework \sysname on Google Cloud. \sysname can also be easily leveraged to characterize the
serving capacity for new CNN models and new hardware combinations. Through
investigating and understanding the model serving performance, we further
identified a number of performance bottlenecks, including inefficient framework
supports for CPU inference and CNN model caching, that hindered the observed
inference performance. Our study forms the basis for complementary research such
as provisioning the inference servers and dispatching inference requests, which
we plan to pursue as the next step.

\section*{Acknowledgment} The authors would like to thank
National Science Foundation grants \#1755659 and \#1815619,
and Google Cloud Platform Research credits.

\balance
\footnotesize{
\bibliographystyle{IEEETran}
\bibliography{perseus.bib}
}
\end{document}